\documentstyle[11pt]{article}  
\newcommand{\be}{\begin{equation}}
\newcommand{\ee}{\end{equation}}
\newcommand{\sg}{\rm{sign}}
\begin{document}
\title{Fermion Mapping for Orthogonal and Symplectic Ensembles}
\author{M. B. Hastings}
\maketitle

\begin{abstract}
The circular orthogonal and circular symplectic ensembles are mapped onto 
free, non-hermitian fermion systems.  As an illustration, the two-level form factors
are calculated.
\today
   \end{abstract}
\section{Introduction}
Dyson introduced the orthogonal, unitary, and symplectic
ensembles of random matrices; different ensembles correspond to
different physical problems
according to the presence or absence of time reversal symmetry.
The joint probability distribution function for the eigenvalues of matrices in
these
ensembles is equivalent to the statistical weight of a configuration of 
charges which repel each other with a logarithmic interaction at an
appropriate temperature.  By integrating over all
levels except for some finite number, level-level correlation functions can
be found\cite{dyson}.  

Although the integrals can be carried out directly, it is
of interest to find a physical system which reproduces the desired 
probability distribution function.  This has previously been done using
the Calogero-Sutherland model\cite{sutherland}, a model of interacting
fermions.  In this paper, physical systems will be presented which use
non-interacting, but non-hermitian, fermion systems to produce the desired 
results.

\section{Circular Ensembles}
The circular orthogonal and symplectic ensembles are defined by considering
a system of $N$ points on the unit circle in the complex plane \cite{dyson}.  
Each 
configuration of points is weighted by a factor $ \prod\limits_{i<j}
|e^{i\theta_i}-e^{i\theta_j}|^{\beta}$, where $\beta=1$ for the orthogonal
ensemble and $\beta=4$ for the symplectic ensemble.  The points are located
at the positions $e^{i \theta_i}$, for $i=1$ to $N$.  Both ensembles will
be mapped onto fermion systems.  

The partition function will be given for
arbitrary chemical potentials, which will permit the calculation of 
correlation functions, reproducing the Dyson formula \cite{dyson2}.  
Although this formula
has been obtained previously by other means, and none of the final
results are new, the fermion mapping for this
problem provides a simple trick for calculating correlation functions.

\section{Orthogonal Ensemble}
First, we will consider the orthogonal ensemble.
We wish to calculate the following partition function 
\be
\label{Z}
\prod\limits_{i=1}^{N} \int d\theta_i \prod\limits_{i<j}
|e^{i\theta_i}-e^{i\theta_j}|
\ee
Then, we will introduce a chemical potential and take
functional derivatives to calculate correlation functions.

Up to a sign, for given $\theta_i$, we may write
\be
\label{e2}
\prod\limits_{i<j} |e^{i\theta_i}-e^{i\theta_j}|
=\pm \prod\limits_{i<j} \frac{e^{i\theta_i}-e^{i \theta_j}}{\sqrt{e^{i\theta_i}
e^{i\theta_j}}}
\ee

We may write eq. (\ref{e2}), up to a constant multiplier in front, as
a correlation function in a two-dimensional free Euclidean fermionic field theory.
We take the $x$ coordinate to be the same as the $\theta$ coordinate,
and thus periodic with period $2\pi$, and take correlation functions to be
analytic in $\tau+ix$.  We introduce the destruction operator $\psi(x,\tau)$ which
destroys a particle at $(x,\tau)$, and the creation operator $\psi^{\dagger}(x,\tau)$.
The field $\psi$ has the action $S=\int dx\,d\tau \psi^{\dagger}(x,\tau)
(\partial_{\tau}+i\partial_{x})
\psi(x,\tau)$.

Then we write
eq. (\ref{e2}) as
\be
\label{fer}
\langle \bigl( \prod\limits_{i=1}^{N} \psi(x_i,0) \bigr) 
\{\psi^{\dagger}(0,-\infty)\}^{N/2}
\{\psi^{\dagger}(0,\infty)\}^{N/2}\rangle
\ee
where the operator $\{\psi^{\dagger}(0,-\infty)\}^{N/2}$ represents 
$N/2$ creation operators, slightly separated from each other in space, and 
located very far
away in time from the line $\tau=0$.  The constant multiplier relating eqs.
(\ref{e2}) and (\ref{fer}) will depend on the precise separation between
the creation points, and the precise distance between those points and $\tau=0$.

We may then forget about the two-dimensional field theory and write eq. 
(\ref{fer}) as the expectation value 
\be
\label{s1}
\langle V^-|\prod\limits_{i=1}^{N} \psi(x_i)|V^+\rangle
\ee
where the state $|V^+\rangle$ is the state obtained by taking the vacuum
state and adding particles to the $N/2$ lowest unoccupied energy states, and
the state $|V^-\rangle$ is the state obtained by taking the vacuum
state and removing particles from the $N/2$ highest occupied energy states.
This expectation value is the correct one since these two states are the
ones projected out by the long time separation used in eq. (\ref{fer}).
It should be noted that the constant factor relating eq. (\ref{s1}) to 
eq. (\ref{e2}) is $N$-dependent, and, unless we take care to determine the
constant, we will only be able to calculate correlation functions, and not
the total free energy.

The sign needed to relate eq. (\ref{e2}) to the product in eq.
(\ref{Z}) may be obtained by introducing an extra field $\eta$.  This
field is $x$-dependent but confined to the line $\tau=0$.  For the $\eta$-field,
$x$ plays the role of time.  This field has the action
$S=\int dx \overline \eta \eta_x$.  Then the sign is simply given
by
\be
\label{s2}
\langle \prod \limits_{i=1}^{N} \{\eta(x_i)+\overline \eta(x_i)\}\rangle
\ee

Then we write introduce a chemical potential $f(x)$, weighting each configuration
of points by $\prod \limits_{i=1}^{N} f(x_i)$, and write the partition function as
\be
\int [d\eta]
\langle V^-|e^{\int dx f(x)\psi(x) (\eta(x)+\overline \eta(x)) 
-\overline \eta \eta_x} |V^+\rangle
\ee
and integrate out $\eta$ to obtain
\be
\label{par}
\langle V^-|e^{\int dx dx' f(x) \psi(x) f(x') \psi(x') \sg(x-x')}|V^+\rangle
\ee
where $\sg(x-x')$ is equal to \be 
\sum \limits_{k} e^{ik(x-x')} \frac{1}{k}
\ee
with $k$ equal to half an odd integer.

It is convenient to rewrite eq. (\ref{par}) for $f(x)=1$ as
\be
\langle V^-|e^{\sum \limits_{k} a(k) a(-k)\frac{1}{k} }|V^+\rangle
\ee
This is clearly equal to 
\be
\label{pr}
(N/2)! \prod \limits_{k=1/2}^{(N-1)/2} 2/k
\ee

We may calculate the two-point correlation function of levels by taking
two derivatives of eq. (\ref{par}) with respect to $f(x)$.  The result 
splits into two pieces, eqs. (\ref{p1}) and (\ref{p2}).

Calculating the correlation function of levels at points $x$ and
$x'$, one piece of the correlation function is obtained by inserting the
operator
\be
\label{p1}
\frac{1}{2}
\psi(x) \psi(x') \sg(x-x')
\ee
into the expectation value of eq. (\ref{par}).  Expanding the $\psi$ operators
in Fourier components $a(k),a(k')$, and noting that the result is 
non-vanishing only if $k=-k'$, the correlation function is
\be
\label{d1}
\frac{1}{2}
\sum \limits_{k=-(N-1)/2}^{(N-1)/2} k 
e^{i k(x-x')} \sum \limits_{l} \frac{1}{l} e^{i l(x-x')}
\ee
where the sum is carried out over {\it all} half odd values of $l$.
The factor of $k$ arises due to a missing factor of $\frac{1}{k}$ in the product
of eq. (\ref{pr}).

The second piece of the correlation function is obtained by inserting
the operator
\be
\label{p2}
\frac{1}{2}
\int dy dy'
\psi(x) \psi(y) \sg(x-y) \psi(x') \psi(y') \sg(x'-y')
\ee
We again expand each $\psi$ operator in $a(k)$ operators, and then pair off
operators $a(k)$ with each other, so that two operators in a pair have
opposite momenta.  If we pair off the first with the second and the third
with the fourth, the result is just the product of one-level correlators.

If we pair off the first operator with the fourth operator, and the
second with the third, we obtain
\be
\label{d2}
\frac{1}{2}
\sum \limits_{k=-(N-1)/2}^{(N-1)/2} \sum \limits_{l=-(N-1)/2}^{(N-1)/2}
e^{i (k+l)(x-x')}
\ee

If we pair off the first operator with the third operator, and the second
with the fourth, we obtain
\be
\label{d3}
\frac{1}{2}
-\sum \limits_{k=-(N-1)/2}^{(N-1)/2} k e^{ik(x-x')} \sum \limits_{l=-(N-1)/2}
^{(N-1)/2} \frac{1}{l} e^{il(x-x')}
\ee

The average spacing between particles is $2\pi/N$.  We rescale the $x$-coordinate
to $Nx$ and rescale the momenta $k$ and $l$ to $k/N$ and $l/N$, to calculate 
the two-level 
correlator in units of average particle spacing.  As $N$ becomes large, we may replace 
the sums by integrals, reproducing the Dyson formula.  We calculate the $m$-th Fourier 
component of the form factor of the two-level correlator, where $m=k+l$.

Then, from equation (\ref{d1}) we obtain
\be
\label{ff1}
-1+|m|\ln(\frac{2|m|+1}{2|m|-1})
\ee
From eqs. (\ref{d2},\ref{d3}) we obtain a non-vanishing result only if
$|m|<1$, in which case eq. (\ref{d2}) gives
\be
1-|m|
\ee
and eq. (\ref{d3}) gives
\be
1-|m|+|m|\ln(2|m|-1)
\ee
So the form factor is given by eq. (\ref{ff1}) for $|m|>1$ and, for
$|m|<1$, is equal to
\be
1-2|m|+|m|\ln(2|m|+1)
\ee

It is also possible to obtain the correlation functions 
of levels which belong to the same alternate series, by introducing
separate chemical potentials for operators $\psi \eta$ and 
$\psi \overline \eta$.  This may help relate this technique to the more standard
method of integration over alternate variables \cite{mehta}.

\section{Symplectic Ensemble}
For the symplectic ensemble, the desired partition function is
\be
\label{Zs}
\prod\limits_{i=1}^{N} \int d\theta_i \prod\limits_{i<j}
|e^{i\theta_i}-e^{i\theta_j}|^4
\ee
For given $\theta_i$, we may again write the weight of the configuration as
a correlation function in a two-dimensional field theory.
Here, the desired correlation function is
\be
\label{fer2}
\langle \bigl( \prod\limits_{i=1}^{N} \psi(x_i,0)^2 \bigr)
\{\psi^{\dagger}(0,-\infty)\}^{N}
\{\psi^{\dagger}(0,\infty)\}^{N}\rangle
\ee
where now $N$ particles must be created at plus and minus infinite time
instead of $N/2$ particles.  The operator $\psi(x_i,0)^2$ represents the
operator 
\be
\lim \limits_{\epsilon\rightarrow 0}\frac{1}{2i\epsilon}\psi(x_i+\epsilon,0) 
\psi(x_i-\epsilon,0)
\ee
This is equal to
\be
\lim \limits_{\epsilon\rightarrow 0}\frac{1}{2i\epsilon}
\sum\limits_{k,k'} e^{i(k+k')x} e^{i(k-k') \epsilon} a(k) a(k')
\ee
which is
\be
\sum \limits_{k,k'} e^{i(k+k')x} \frac{k-k'}{2} a(k) a(k')
\ee

Taking $|V^+\rangle$ to be the state with $N$ particles added to the vacuum, and
$|V^-\rangle$ to be the state with $N$ particles removed from the vacuum, 
eq. (\ref{Zs}) becomes, up $N$-dependent factors,
\be
\label{sf}
\langle V^-|e^{\int dx f(x) \psi(x)^2}|V^+ \rangle
\ee
where a chemical potential $f(x)$ has been introduced.  For $f(x)=1$, this
is equal to
\be
\label{y2}
N! \prod \limits_{k=1/2}^{2N-1/2} 2k
\ee

The calculation of the two-level form factor at this point is almost identical
to that in the orthogonal ensemble.  We must introduce the operator 
$\frac{1}{2}\psi(x)^2 \psi(x')^2$ into the expectation value, expand in Fourier 
components, and pair off the annihilation operators in all possible ways.  

The non-trivial pairings of annihilation operators involve pairing the first
with the third or the first with the fourth.  These two possibilities give the
same contribution.  This yields
\be
\label{y4}
-\frac{1}{4} e^{i (k+l) (x-x')} \frac{(k-l)^2}{kl}
\ee
which is zero for $|m|>2$ and equal to, in the large $N$ limit,
\be
2-|m|+\frac{|m|}{2}ln(|m|-1)
\ee
for $|m|<2$.
The factor of $1/4$ in eq. (\ref{y4}) arises from the the factor of $1/2$ in the
operator inserted into the expectation value, times the factor of $2$ from
the two different pairings, times the factor of $1/4$ from the two terms missing
in eq. (\ref{y2}).

A fermion mapping has been given for the orthogonal and symplectic ensembles.
Although closely related to the technique of integrating over alternate
variables, this technique provides a physical system, the fermionic field
theory, which handles the various integrals and determinants.

\end{document}